# Tailoring the Spectral and Directional Emissivity of Functionalized Laser Processed Surfaces

ANDREW BUTLER[1], ANDREW REICKS[1], DENNIS ALEXANDER[1], GEORGE GOGOS[2], CRAIG ZUHLKE[1], AND CHRISTOS ARGYROPOULOS[3,*]

[1]*Department of Electrical and Computer Engineering, University of Nebraska-Lincoln, Lincoln, NE 68588, USA*

[2]*Department of Mechanical and Materials Engineering, University of Nebraska-Lincoln, Lincoln, NE 68588, USA*

[3] *Department of Electrical Engineering, The Pennsylvania State University, University Park, PA 16803, USA*

*\*cfa5361@psu.edu*

**Abstract:** Development of methods to control the directional and spectral characteristics of thermal radiation from metallic surfaces is a critical factor enabling many important thermal management applications. In this paper, we study the thermal emission properties of functionalized aluminum surfaces produced through femtosecond laser surface processing (FLSP). These types of surfaces have recently been found to exhibit near-unity broadband omnidirectional emissivity. However, their ultrabroadband absorption response includes visible and near-infrared (IR) radiation, in addition to the mid-IR range, which limits their use as daytime passive radiative cooling devices. Here, we present ways to solve this problem by demonstrating a new design that uses a dielectric Bragg visible light reflector to accurately control the thermal emission spectra of the FLSP surface with the goal of achieving high performance daytime radiative cooling operation. In addition, we propose other designs based on dielectric multilayer structures to further tailor and control the spectra and thermal emission angles of the FLSP surfaces leading to narrowband and broadband directional thermal radiation. The presented photonic engineering approach combined with FLSP structures will be beneficial to various emerging applications, such as radiative cooling, thermal sensing, and thermophotovoltaics.

## 1. Introduction

Thermal radiation is a natural phenomenon with many important applications including thermal sensing and imaging [1], thermophotovoltaics [2–5], and radiative cooling [6]. The thermal emissivity of material surfaces is primarily characterized by its spectral response and angular direction of thermal radiation. Engineering the emissivity characteristics of material surfaces is challenging but of paramount importance since it promises to enable many new and intriguing thermal management technologies. With recent improvements in photonic engineering, technology to control the spectral emissivity of photonic structures has rapidly evolved, leading to various emerging applications such as daytime passive radiative cooling and thermophotovoltaics [2–5]. However, a more challenging and somewhat less explored aspect of thermal radiation manipulation is relevant to the photonic control over the direction combined with the spectrum of thermal emission from a surface. The spatial and frequency control of thermal emission has the potential to make great improvements to many thermal management applications by reducing heat loss or absorption at oblique angles or different wavelength ranges leading to improved efficiency and performance of thermal devices.

More specifically, passive radiative cooling is currently one of the main applications where efficient spectral emissivity control is required. Interestingly, a variety of techniques for designing surfaces and structures with special spectral emissivity characteristics exist in the literature [7–17]. More specifically, in order to achieve daytime passive radiative cooling, a structure needs low emissivity in the solar wavelength range (300 nm – 4 µm) to avoid absorbing heat from the sun and high emissivity in the mid-IR wavelengths (8 µm – 13 µm), i.e., the so-called atmospheric window, where the atmosphere is transparent, and the thermal emission can be radiated out into space. One recent design approach is to use nanoparticle embedded paints, coatings, and polymers to achieve high solar reflection, i.e., low absorption, and high mid-IR emissivity …21] [10,18–24]. In particular, one noteworthy design used a glass-polymer hybrid metamaterial which incorporated silicon oxide ($SiO_2$) microspheres that thermally emit highly in the atmospheric window and can be quickly produced using roll-to-roll manufacturing [19]. More recently, substantial daytime cooling of 1.7 °C below ambient temperature was demonstrated using an acrylic paint that incorporated multi-sized calcium carbonate ($CaCO_3$) nanoparticles which provided high emission in the atmospheric window while scattering and reflecting 95.5% of solar light [18]. Many polymer materials suffer from degradation under exposure to ultraviolet light. To this end, other recent designs have focused on using ceramic materials or glass to improve durability [25,26]. Durability is a main concern for realistic implementations of radiative cooling [27].

One very common approach to meeting the stringent criteria for daytime passive radiative cooling is by using nanophotonic inspired one-dimensional (1D) multilayer structures [11,12,15,28–32]. Typically, these designs use a metallic back reflector with a dielectric-based multilayer thermal emitter stacked on top of it. The metallic reflector provides high solar reflectance which may be further boosted by the dielectric layers on top. Materials that naturally absorb in the mid-IR range are chosen to constitute the layered nanophotonic emitter structures leading to high emissivity in the atmospheric window. Two-dimensional (2D) photonic cooling devices operate on a similar principle but with more complex nanostructures included in their design that mainly aid to tune the desired emission and reflection responses in different wavelength ranges [13,33–35].

On a relevant note, the spatial control and confinement of the thermal emission pattern is also interesting for some applications where directional thermal emitters are required, such as thermophotovoltaics [2,3]. Such directional emitters can have narrowband or broadband response depending on the application. Broadband directional thermal radiation is more challenging to be realized. One approach to accomplish directional control in thermal emission is to use epsilon near zero (ENZ) materials [36]. While these materials show preferential thermal emission at larger angles, their effect is inherently narrowband. Recently it was demonstrated that stacking multiple layers of different ENZ materials could result in broadband directional emission, however, the emissivity was relatively low and worked only for transverse magnetic (TM) polarization [37]. Thermal emission is incoherent radiation composed of both TM and transverse electric (TE) polarizations; thus, the aforementioned configuration will be limited for practical applications. Another theoretical approach was proposed using hyperbolic materials, though this design would be difficult to realize experimentally [38]. Other relevant configurations rely on resonances that are inherently narrowband in nature [39,40] or require more complicated nanostructures in their design such as metallic gratings or nanorods [41–43]. Polarization independent directional optical filtering has been demonstrated using a multilayer dielectric stack [44]. The effect is narrowband and restricts the transmission of light to near normal incidence. A broadband directional optical filter was also experimentally realized that utilized Brewster angle transmission and an impedance matched external medium to accomplish a broad and highly directional transmission spectra [45]. Since it utilizes the Brewster angle, this effect works only for TM polarization though.

In our current work, we theoretically demonstrate that femtosecond laser surface processed (FLSP) surfaces can operate as near perfect omnidirectional absorption or emission substrates [46] with absorption properties extended from mid-IR to visible wavelengths. While such broad absorption spectrum or thermal emission response is ideal for nocturnal radiative cooling devices, it needs to be controlled in an effective way to apply FLSP technology to the counterpart emerging application, i.e., daytime passive radiative cooling. To solve the FLSP broad absorption problem, we designed three dielectric multilayer coatings that control the spectral and directional emissivity of the FLSP structure. The first design achieves high performance daytime passive radiative cooling by reflecting incident sunlight and allowing the FLSP surface to thermally radiate through the entire mid-IR atmospheric window wavelength range. The second design demonstrates narrowband directional thermal emission, effectively turning the FLSP composite structure into a narrowband thermal emission antenna. The final design utilizes Brewster angle transmission to achieve broadband directional thermal emission limited to TM polarization. Note that the emissivity of the FLSP surface is significantly higher and more omnidirectional than just simple coating a metallic mirror with alumina, as was demonstrated in our previous work [46].

The current work is an initial step in applying the unique characteristics of FLSP surfaces and the FLSP technique to thermal radiation applications. FLSP is a scalable process that can be used to quickly functionalize objects with large surface areas. The FLSP technique can be applied directly to devices with metallic exteriors, making the object itself radiate thermal energy rather than needing heat to propagate through multiple layers. This work demonstrates that the thermal emission of FLSP microstructures can be successfully coupled with photonic structures to control its spectral and angular properties. All presented designs can be used in various emerging thermal management applications.

## 2. Results

### 2.1 FLSP Broadband Absorption

We begin our investigation by demonstrating that the FLSP technology based on aluminum (Al) surfaces leads to ultrabroadband absorption that spans visible and IR wavelengths. FLSP is an advanced manufacturing technique that can produce a wide range of surface structures with unique optical properties. By controlling laser parameters such as the number of pulses, pulse duration, and laser fluence many different types of periodic or quasi-periodic structures can be produced usually on metallic surfaces using a femtosecond laser system [47-50]. We have previously presented experimental results from FLSP on an Al surface that produced mounds of Al covered by a thick layer of aluminum oxide ($Al_2O_3$) [46]. The absorptance of the aluminum FLSP surface structure is near perfect in the mid-IR, as was theoretically verified and experimentally measured in our previous work [46].

Here, we extend our theoretical calculations in the visible and near-IR range by simulating such Al-based FLSP metallic surface using the Finite Element Method (FEM) (COMSOL Multiphysics®). The FLSP surface is modelled as periodic two-dimensional (2D) concentric half-circles with a unit cell simulation schematic shown in Fig. 1a. We perform 2D simulations since they produce identical results with three-dimensional (3D) simulations but are much faster to be executed and require much less computational memory. The inner circle is composed of Al with a radius of 15 μm and the outer half circle is made of $Al_2O_3$ with a longer radius of 30 μm. A thin 300 nm thick layer of Al lies underneath the half-circles to make the total structure opaque to incident electromagnetic radiation. The realistic (i.e., frequency dispersive) dielectric constants of metal (Al) [51] and dielectric ($Al_2O_3$) [52,53] are used to simulate the materials. Periodic boundary conditions are applied to both left and right sides of the 2D simulation domain shown in Fig. 1a.

The spectral and angular emissivity of this structure is calculated from the S-parameters derived by the simulations. More specifically, the structure's absorptance $A$ can be computed by the formula: $A=1-|S_{11}|^2-|S_{21}|^2$, where $S_{11}$ and $S_{21}$ are the reflection and transmission coefficients, respectively. By using the Kirchhoff's law of thermal radiation [17], the computed absorptance equals the emissivity of any structure in thermal equilibrium, similar to the current FLSP design. Since the polarization of thermal radiation is random, each simulation is run once for TM polarization and again for TE polarization. The derived emissivity results are computed by the average value of the two polarizations. The computed angular absorptance of the simulated FLSP structure at a fixed mid-IR wavelength of 10 μm is demonstrated in Fig. 1b. Omnidirectional performance is obtained similar to the mid-IR experimental results presented in our previous work [46]. The computed absorption spectrum spanning mid-IR, near-IR, and visible wavelengths is shown in Fig. 1c, where notable the absorption is very high in both mid-IR, as was presented in our previous work [46], but also in the near-IR/visible spectra. The high visible absorption is the reason why FLSP structures usually turn black after intense laser illumination [54]. Our theoretical results presented in Fig. 1c clearly demonstrate that this type of metallic FLSP structure possesses near ideal emissivity in the atmospheric window (blue highlighted area in Fig. 1c), which is exactly what is required for radiative cooling applications. However, its absorption in the solar spectrum is also very high, as clearly depicted in the zoomed in plot of Fig. 1c, making such FLSP structure incapable of radiative cooling, since it will heat up under direct solar light. However, the currently presented FLSP structure will indeed achieve nocturnal radiative cooling or even daytime cooling if the structure is not under direct sunlight, i.e., it is placed in a shadowed region.

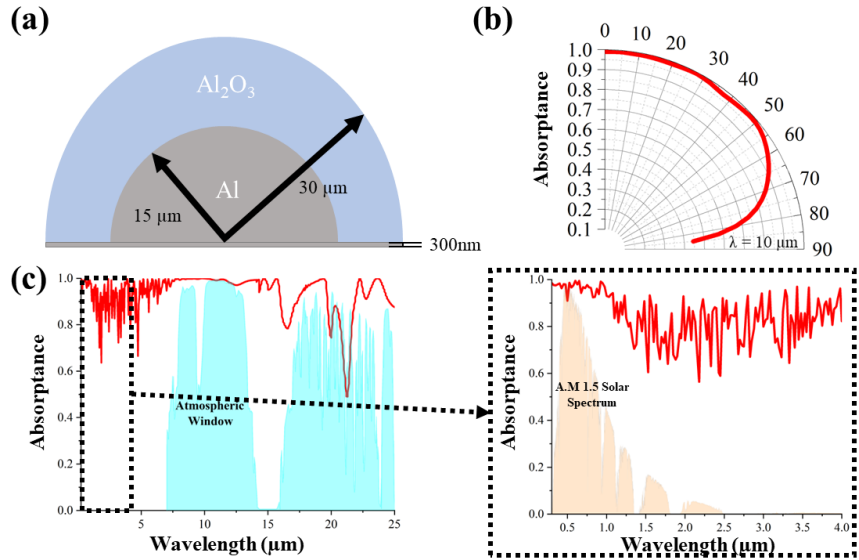

Fig 1. a) Schematic of the Al-based FLSP structure 2D model used in our simulations. b) Angular absorptance of the FLSP structure computed at a fixed mid-IR wavelength of 10μm. c) Simulation results of the FLSP absorptance spectra when the structure is illuminated at normal incidence. The atmospheric window emission spectrum is also plotted as the blue highlighted area and the solar spectrum is shown in the zoomed in plot.

*2.2 FLSP Daytime Passive Radiative Cooling*

While the FLSP structure almost fully absorbs or emits thermal radiation at the atmospheric window, as demonstrated in Fig. 1c, it exhibits similar performance in visible wavelengths which limits its applicability as daytime radiative cooling device. Hence, to prevent the FLSP structure from absorbing too much solar radiation, we designed a new specialized Bragg mirror optical filter that reflects solar light but allows mid-IR radiation to pass through. The proposed distributed Bragg visible light reflector sits on top of the FLSP mounds with relevant schematic demonstrated in Fig. 2. It is made of alternating layers of high and low refractive index materials that are feasible to experimentally realize. The center wavelength of the presented optical reflector, acting as stopband optical filter in our design, can be adjusted by using quarter-wavelength layer thicknesses. The total thickness of the proposed stopband filter is determined by the difference between the refractive indices of the alternating materials.

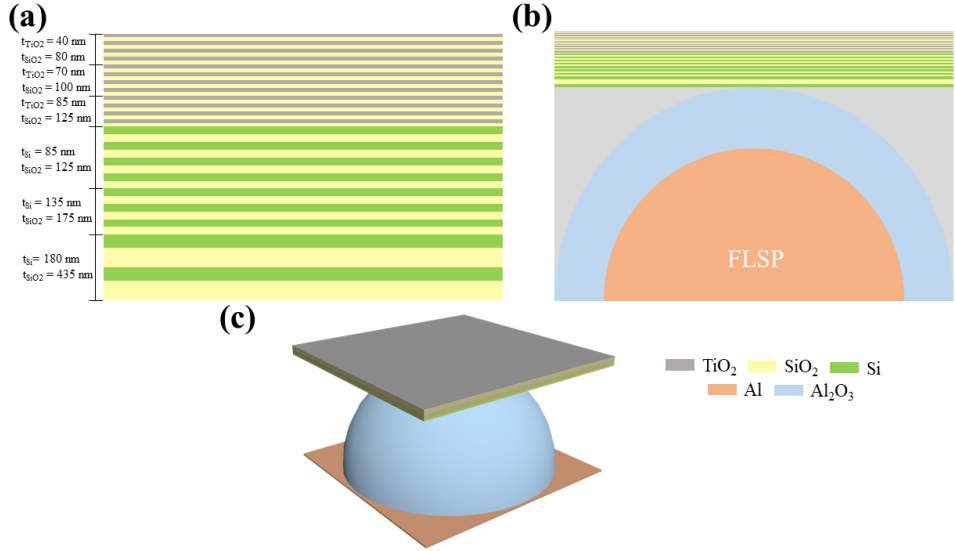

Fig. 2. a) The designed Bragg visible light reflector where each layer thickness value is also demonstrated. b) 2D schematic of envisioned daytime passive radiative cooling FLSP device. c) 3D schematic of FLSP cooling device along with details of all materials used.

According to the AM1.5 Global Tilt spectra [55], on a typical day there is an average of ~1000 W/m$^2$ of irradiance from the sun. In a realistic scenario, peak intensity is reached at noon but can vary depending on weather conditions. With such a large amount of incident power, even 5% absorption in the solar spectrum results in ~50 W/m$^2$ of intensity being absorbed which induces substantial heating that has a detrimental impact on the performance of the cooling device during the daytime. Therefore, it is extremely important that the reflectance of the designed distributed Bragg mirror will be as close to 100% as possible over the entire solar spectrum leading to zero absorption in the same range.

The currently designed visible reflector consists of four different materials and several groups of layers with varying thicknesses, where their dimensions are depicted in Fig. 2a. The top layers consist of titanium dioxide ($TiO_2$) as a high refractive index dielectric material and silicon dioxide ($SiO_2$) as a low index material. There are three unit cells composed of one $TiO_2$ layer and one $SiO_2$ layer with fixed thicknesses, where each group of three unit cells is repeated four times, leading to a total of twenty four individual material layers (either $TiO_2$ or $SiO_2$). The bottom layers of the entire distributed Bragg reflector structure consist of silicon (Si) as the high index material and $SiO_2$. There are three unit cells composed of one Si layer and one $SiO_2$ layer with fixed thicknesses. The first unit cell group is repeated four times, while the second unit cell group is repeated three times and the last unit cell arrangement is repeated twice, leading to a total of eighteen individual material layers (either Si or $SiO_2$).

To compute the reflectance and, as a result, absorptivity/emissivity of the proposed visible light reflector, we developed an analytical approach based on the modelling of each reflector layer as a finite transmission line and using their ABCD parameters to calculate the S-parameters of the entire structure [56]. We also simulated the entire reflector filter by using COMSOL Multiphysics® to verify our analytical model. Interestingly, both analytical (theory) model and numerical simulations produce very similar results. The ABCD matrix of an arbitrary layer numbered m in the reflector design is given by:

$$\begin{bmatrix} A & B \\ C & D \end{bmatrix}_m = \begin{bmatrix} \cos(kt) & jZ\sin(kt) \\ j\frac{1}{Z}\sin(kt) & \cos(kt) \end{bmatrix}, \quad (1)$$

where $Z = \eta_0\sqrt{\epsilon - \sin^2(\theta)}$ for oblique TM polarization and $Z = \eta_0/\sqrt{\epsilon - \sin^2(\theta)}$ for oblique TE polarization is the effective impedance of each layer material, $\eta_0$ is the impedance of free space ($\eta_0 \sim 377\ \Omega$), $t$ is each layer thickness, $k = k_0\sqrt{\epsilon - \sin^2(\theta)}$ is the wavevector in each layer material, $\varepsilon$ is the frequency dependent complex dielectric constant of each layer material, and $\theta$ is the angle of incidence. The ABCD matrix of the entire reflector can be calculated by chaining together the ABCD matrices of each layer, where the total number of layers is forty-two:

$$\begin{bmatrix} A & B \\ C & D \end{bmatrix}_{tot} = \begin{bmatrix} A & B \\ C & D \end{bmatrix}_1 \begin{bmatrix} A & B \\ C & D \end{bmatrix}_2 \cdots \begin{bmatrix} A & B \\ C & D \end{bmatrix}_{42}. \quad (2)$$

The reflectance and absorptance of the reflector can then be obtained by converting the ABCD parameters of the reflector to its S-parameters [56]:

$$S_{11} = \frac{AZ_0 + B - CZ_0^2 - DZ_0}{AZ_0 + B + CZ_0^2 + DZ_0}, \quad (3)$$

$$S_{21} = \frac{2}{A + \frac{B}{Z_0} + CZ_0 + D}, \quad (4)$$

where the free space impedance is $Z_0 = \eta_0\cos(\theta)$ for oblique TM polarization and $Z_0 = \eta_0/\cos(\theta)$ for oblique TE polarization. The total reflectance is equal to $|S_{11}|^2$, while the transmittance is equal to $|S_{21}|^2$. Finally, the computed absorptance ($A$) and emittance ($E$) are equal to $E = A = 1 - |S_{11}|^2 - |S_{21}|^2$ due to Kirchhoff's law of thermal radiation for objects in thermal equilibrium [17].

The reflectance and transmittance spectra of the designed optical Bragg reflector filter are shown in Figs. 3a and 3b, respectively. Note that these results are just for the filter, i.e., no FLSP included in the model. The frequency dependent (dispersive) complex dielectric constants of $TiO_2$ [57], $SiO_2$ [58,59], and Si [60,61] are used to accurately model each material both in theory and simulations. For wavelengths above the range of available data (>14 µm for Si), the dielectric constants were extrapolated as constants. The results of the analytical model (red lines Figs. 3a and 3b) were found to match very well the relevant FEM simulation results (blue lines in Figs. 3a and 3b). While the results in blue lines in Figs. 3a and 3b are for normal incidence, it is interesting to note that the reflector also works in a similar way for oblique incident angles, as demonstrated later in Fig. 4b (blue line). In addition, the reflector has the same response for both TE and TM polarization, so it is ideal to emit incoherent thermal radiation. Figure 3c shows the real and imaginary parts of the frequency dependent refractive indices used in these calculations [57 – 61]. While oxide materials are highly absorptive in the infrared range, each oxide has different wavelength ranges where they absorb more strongly. The high transmittance ranges of our Bragg reflector filter are due to gaps between the absorption spectra of $SiO_2$ and $TiO_2$.

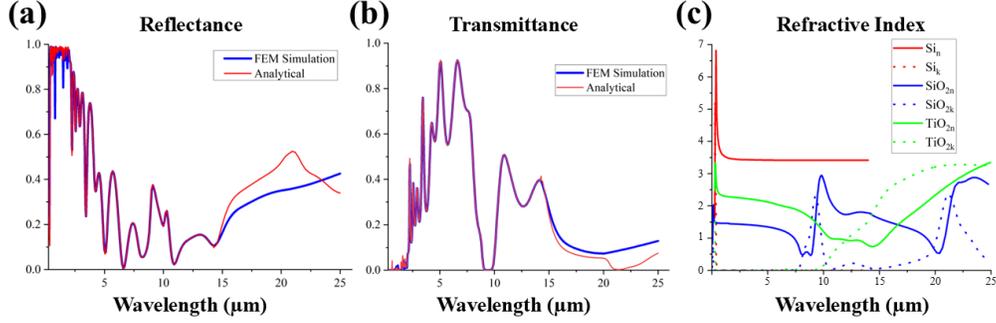

Fig. 3. Numerical simulation and analytical modeling results for the a) reflectance and b) transmittance spectra of the designed optical Bragg reflector filter at normal incidence. c) Real (n) and imaginary (k) parts of the refractive indices used in the calculations [57-61].

The stopband filter operation wavelength range of each pair of layers depends on the difference between the refractive indices of their materials. As an example, Si has a much larger refractive index value than $TiO_2$, but Si also has notable losses in ultraviolet (UV) wavelengths including part of visible range (<0.5μm). Interestingly, $TiO_2$ also has some absorption in the UV frequencies, though much less compared to Si, while $SiO_2$ is lossless over the entire solar spectrum including part of UV. The top layers of $TiO_2$ lead to high reflectance in the small wavelengths of the visible spectrum. Many layers of $TiO_2$ are needed because the refractive indices of $TiO_2$ and $SiO_2$ have relatively close values which reduce their reflectance. The middle Si layers provide high reflectance for the longer wavelengths in the near-IR range. The obtained large reflectance values in the wavelength range between the end of the solar spectrum (4 μm) and the beginning of atmospheric window (8 μm) help reduce the absorption of the atmosphere's inherent thermal radiation. The presented distributed Bragg reflector design achieves an average reflectance of 90.1% between 0.3 μm and 2.5 μm at normal and oblique incident angles. Materials such as $TiO_2$ and $SiO_2$ are commonly used in radiative cooling devices because they emit strongly in the mid-IR at wavelengths within the atmospheric window. By simultaneously achieving high solar reflection and mid-IR thermal emission, the distributed Bragg reflector used in our cooling device is actually capable of daytime radiative cooling on its own. However, this cooling performance is notably increased when combined with the FLSP emitter used as its substrate which is demonstrated in the following.

To calculate the actual cooling power of the designed Bragg reflector, we first consider the power radiated from the device ($P_{rad}$). This is the power that contributes to the overall cooling performance of the device. We then need to consider sources of heat that counteract this radiative cooling effect. Typically, these are incident power from the sun ($P_{sun}$), inherent thermal emission from the atmosphere ($P_{atm}$), and conductive and convective heat sources from the environment ($P_{cdcv}$). The net cooling power $P_{cool}$ of the device is then defined as:

$$P_{cool} = P_{rad} - P_{sun} - P_{atm} - P_{cdcv}. \qquad (5)$$

The power radiated from the device is given by:

$$P_{rad}(T) = A \int_0^{\frac{\pi}{2}} \sin(\theta) \cos(\theta)\, d\theta \int_0^\infty I_{BB}(T,\lambda)\varepsilon(\lambda,\theta) d\lambda, \qquad (6)$$

where $A$ is the area of the cooling device (not to be confused with absorptance), $T$ is the temperature of the cooling device, $\varepsilon(\lambda, \theta)$ is the spectral and angular emissivity of the cooling device, and $I_{BB}(T, \lambda)$ is the blackbody spectral radiance given by Planck's Law: $I_{BB}(T,\lambda) =$

$\frac{2hc^2}{\lambda^5} \frac{1}{e^{hc/(\lambda k_B T)}-1}$, where $h$ is Planck's constant, $k_B$ is Boltzmann's constant, $c$ is the speed of light, and $T$ is the temperature. The power absorbed due to the inherent thermal emission of the atmosphere is given by:

$$P_{atm}(T_{amb}) = A \int_0^{\frac{\pi}{2}} \sin(\theta) \cos(\theta)\, d\theta \int_0^{\infty} I_{BB}(T_{amb}, \lambda) \varepsilon_{atm}(\lambda, \theta) \varepsilon(\lambda, \theta) d\lambda, \quad (7)$$

where $T_{amb}$ is the ambient temperature of the air and $\varepsilon_{atm}(\lambda, \theta)$ is the spectral and angular emissivity of the atmosphere given by: $\varepsilon_{atm}(\lambda, \theta) = 1 - t(\lambda)^{\frac{1}{\cos(\theta)}}$ [62]. For our calculations, we use the atmospheric transmission spectra $t(\lambda)$ of the sky in Mauna Kea, Hawaii with 1 mm of precipitable water vapor [63–65] and an ambient temperature $T_{amb}$ of 300K. Note that this represents near ideal atmospheric conditions. Higher levels of precipitable water vapor will reduce the transmittance through the atmospheric window and decrease the cooling power. The power absorbed by the cooling device due to solar radiation is given by:

$$P_{sun} = A \int_0^{\infty} \varepsilon(\lambda, \theta_{sun}) I_{AM1.5}(\lambda) d\lambda, \quad (8)$$

where $I_{AM1.5}(\lambda)$ is the spectral irradiance of the AM1.5 Global Tilt spectrum [55] and $\theta_{sun}$ is the angle of the sun with respect to the surface of the cooling device. For our calculations, we assume that the sun is normally incident upon the surface of the cooling device, which is the worst-case scenario in terms of heating. Finally, the power absorbed by the cooling device due to conductive or convective heat sources is given by:

$$P_{cdcv} = A h_{cdcv}(T_{amb} - T), \quad (9)$$

where $h_{cdcv}$ is the heat transfer coefficient that accounts for heating due to both conduction and convection. For our calculations, we assume a heat transfer coefficient of 10 W/m²K, which is a typical value used in previous works [66].

The computed absorptance of the designed optical Bragg reflector filter is shown with the blue line in Fig. 4a, while a polar plot of its angular absorptance at a fixed wavelength of $\lambda$ = 11 µm is shown in Fig. 4b (again blue line). These plots clearly demonstrate that the absorption or emission of the filter is omnidirectional, similar to the FLSP structure response. However, the filter's absorption is relatively high at mid-IR (again similar to FLSP but lower values) while being much lower at visible wavelengths compared to FLSP structure response. These are ideal properties to achieve daytime passive radiative cooling performance.

Next, using Eq. (1), we calculate a cooling power of 61.77 W/m² at 300K and an equilibrium temperature of 295 K for the distributed Bragg reflector (no FLSP yet), as shown in Fig. 4c by the blue line. The computed equilibrium temperature of the structure is plotted in Fig. 4d (blue line) as a function of time over the course of a typical day with variations in the ambient temperature demonstrated by the black line in Fig. 4d. The sun is assumed to rise at hour 6 and set at hour 21, similar to a typical summer day. Incident solar radiation is only included in the calculations between these hours. This plot serves to demonstrate the cooling performance of the bare filter structure in a realistic scenario. Indeed, it is proven that the distributed Bragg reflector is capable of radiative cooling on its own. However, the addition of the FLSP Al-based surface below this optical filter further increases the power emitted in the mid-IR atmospheric window because the FLSP surface absorbs any remaining radiation transmitted through the filter. To obtain the emissions spectra of the FLSP surface combined with the Bragg reflector, we combine them and simulate the entire structure in 2D, as depicted in Fig. 2c.

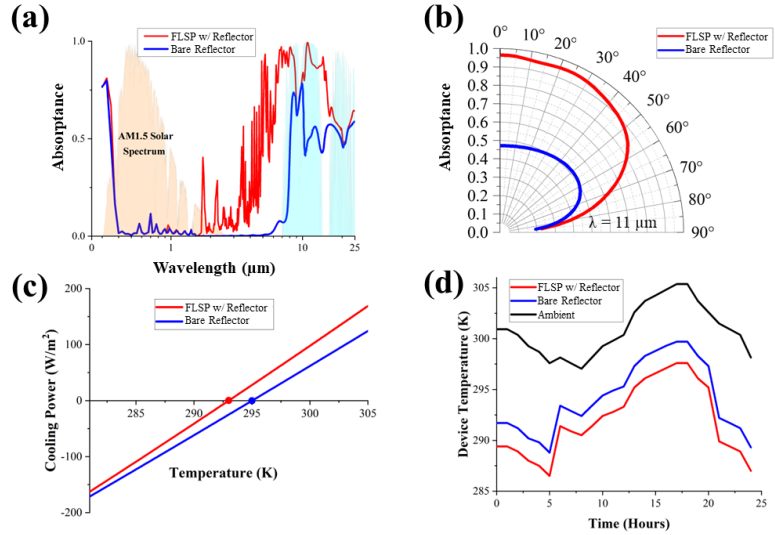

Fig. 4. a) Computed absorptance spectra. b) Angular absorptance at a fixed wavelength of λ = 11 µm. c) Calculated cooling power vs temperature. d) Equilibrium temperature over the course of a day. In all figures, the blue lines correspond to the bare Bragg reflector while the red lines depict the results of the reflector combined with the FLSP Al-based substrate. The ambient temperature, shown as black line in (d), is the temperature measured hourly during a typical summer day.

The absorptance spectra of the composite FLSP Al-based substrate plus reflector cooling device are shown in Fig. 4a with the red line. The angular absorptance at a fixed wavelength of λ = 11 µm is also depicted by the red line in Fig. 4b, proving that the structure still has an omnidirectional response. Note that the omnidirectional absorptance is substantially higher for the composite FLSP structure (red line/Fig. 4b) compared to just the reflector (blue line/Fig. 4b) at this specific wavelength. The FLSP cooling device achieves a cooling power of 97.79 W/m$^2$ at 300K and reaches an equilibrium temperature of 293K, as depicted in Fig. 4c. While the reflector provides very high reflectance in the solar spectrum, the absorptance of the FLSP device is very high across much of the mid-IR spectrum, including outside of the atmospheric transmission windows (red line in Fig. 4a). At wavelengths where the atmospheric transmission is low, this reduces the cooling power since extra power is absorbed from the inherent thermal radiation of the atmosphere. However, the total radiated power in the atmospheric window is much higher with the FLSP composite device compared to the bare reflector alone, which is the reason why it performs better in terms of daytime passive radiative cooling.

While the proposed FLSP cooling device can perform well under standard conditions and achieve improved cooling than the bare reflector (see comparison in Fig. 4d), the extra absorption of atmospheric emission slightly degrades its overall cooling performance. Note that there are applications where increased absorption outside of the atmospheric window is beneficial. For example, when the temperature of the cooling device is higher than the ambient temperature, the power absorbed from the atmosphere will be less than the power radiated by the device. This is because the blackbody radiance used in Eq. (6) varies with the temperature of the device, while the blackbody radiance in Eq. (7) uses a constant ambient temperature. The black body radiance increases as the temperature increases, hence, above ambient temperature, $P_{rad}$ is greater than $P_{atm}$. This makes the presented FLSP Al-based cooling device ideal for above ambient temperature cooling applications such as the cooling of industrial buildings or equipment that are prone to heating. It is also very suitable for high altitude applications such as aircraft or spacecraft coatings where the atmosphere is much thinner or of no concern, respectively.

It is worth mentioning that the presented FLSP technology could potentially be well applied to any device with a metallic exterior. The versatility and scalability of the FLSP process provides great potential for a wide range of radiative cooling scenarios. In most applications, the radiative cooling device is being used to cool another nearby object such as a building. An additional benefit provided by our FLSP design is that the thermal emitter is put in direct contact with the desired object. The heat applied to the cooling device by the object is transferred directly to the emitter rather than allowing it to propagate through several layers of intermediate materials.

*2.3 Narrowband Directional Thermal Emission*

The design above utilizes the near omnidirectional thermal emission of the FLSP surface to achieve high performance radiative cooling. However, confining thermal emission to narrow angles is also of interest and could have applications in the design of emerging infrared devices such as detectors and directional thermal emitters. Similar to the above, multilayer stack optical filter designs can be used to restrict not only the spectral band of thermal emission, but also the direction of resulted thermal radiation. Figure 5a shows the schematic of a simple multilayer structure that can be used as an optical spectral and directional filter to restrict the thermal emission of the FLSP surface to angles near normal to the surface confined in narrow wavelength bands. This filter design is polarization independent and utilizes the edge of its stopband region to restrict the thermal emission to near-normal angles [44]. In this case, Si is used as the high index material and calcium fluoride ($CaF_2$) is used as the low index material. Interestingly, $CaF_2$ is utilized instead of $SiO_2$ in this design because it has a similarly low refractive index but is less lossy than $SiO_2$ in the mid-IR range of interest. The $CaF_2$ layers were accurately simulated using the $CaF_2$ frequency dependent permittivity values [67]. The extinction coefficient of $CaF_2$ is very small (assumed to be negligible) in the wavelength range considered, so we used only the real part of the dielectric constant in our simulations.

Figure 5b demonstrates the emissivity of the designed multilayer structure placed on top of the FLSP surface for both TM (left plot) and TE (right plot) polarizations obtained through FEM simulations. Figure 5c represents the calculated angular absorptance pattern of the composite structure at a fixed wavelength of $\lambda = 7.83$ μm, where the peak emission of the structure is also achieved. Figure 5d exhibits the computed transmittance of only the multilayer structure (no FLSP) for both TM (left plot) and TE (right plot) polarizations obtained through the analytical transfer matrix method described above. The results are very similar to the absorption of the composite structure (FLSP plus filter design) depicted in Fig. 5b.

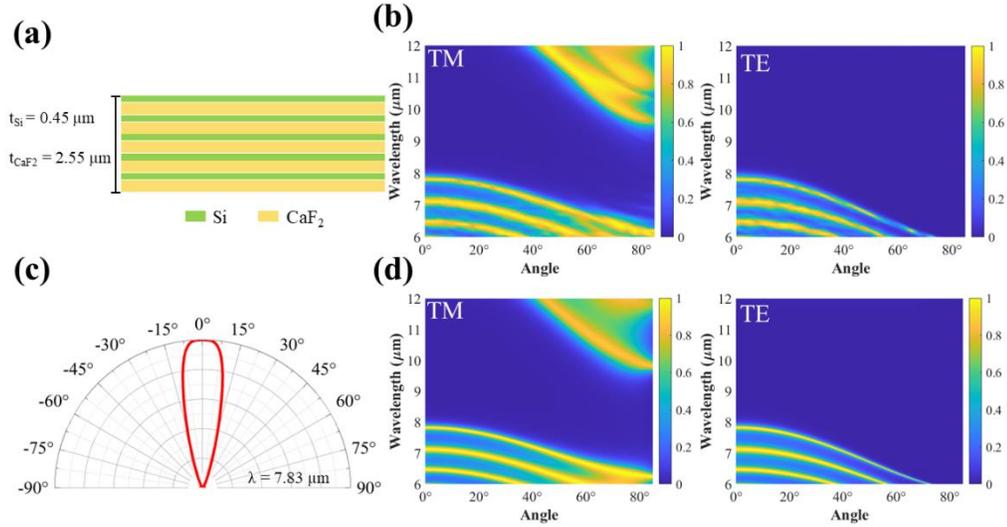

Fig. 5. a) Schematic of the multilayer filter structure used for narrowband directional thermal emission. b) Absorptance spectra of the FLSP surface covered by the multilayer structure for TM (left) and TE (right) polarizations. These results were obtained through FEM simulations. c) Polar plot of the filter's angular absorptance at a fixed wavelength of λ = 7.83 µm. d) Transmittance spectra of the bare multilayer filter structure (i.e., without FLSP) for TM (left) and TE (right) polarization. These results were obtained analytically using the transfer matrix method.

For the layer thicknesses with values shown in Fig. 5a, the edge of the designed filter's stopband region occurs at a wavelength of 7.83 µm. At this mid-IR wavelength, the high emissivity is restricted to angles within about plus/minus 15° from normal, as demonstrated in Figs. 5b-d. The fringe higher-order resonances appearing at lower wavelengths have a similar effect though they also exhibit some emissivity at higher angles. The edge of the stopband, where directional emission is obtained, can be easily tuned by varying the thickness of each material layer. Thus, this type of multilayer optical filter structure can be used for a variety of applications where narrowband coherent IR directional light source operation is required [68].

2.4 Broadband Directional Thermal Emission

It is also possible to restrict the broadband thermal emission to a narrow range of angles, different from the previous section design where the optical filter had narrowband response operating directionally only at 7.83 µm. This type of multilayer filter structure utilizes impedance matching with the surrounding medium to achieve broadband thermal emission only at the Brewster angle [38]. The schematic of this multilayer structure is shown in Fig. 6a. Three multilayer stacks with different periodicities are used to realize the broadband effect. Each stack consists of eight layers of alternating Si and $CaF_2$ materials. For this particular design, the multilayer structure is placed on top of the FLSP and the area above the multilayer filter is filled with a material that is impedance matched to $CaF_2$. To model this structure analytically, Eqs. (1)-(4) given before were simply adapted by dividing $\eta_0$, $\varepsilon$, and Z by the dielectric constant of the surrounding medium, $\varepsilon_m$.

For our simulations and analytical calculations, we used an external surrounding medium with a refractive index of 1.35 or $\varepsilon_m = 1.8225$, which is similar value to the $CaF_2$ permittivity [67]. The computed absorptance of the multilayer structure combined with the FLSP surface is shown in Fig. 6b for TM (left plot) and TE (right plot) polarizations. In addition, the polar plot of the absorptance of the same composite structure as a function of the incident angle for a fixed wavelength of λ = 7.8 µm is shown in Fig 6c, where the directional response is clearly demonstrated. The transmittance of the multilayer structure and impedance matching surrounding medium without the FLSP below the optical filter is shown in Fig. 6d for TM (left plot) and TE (right plot) polarizations.

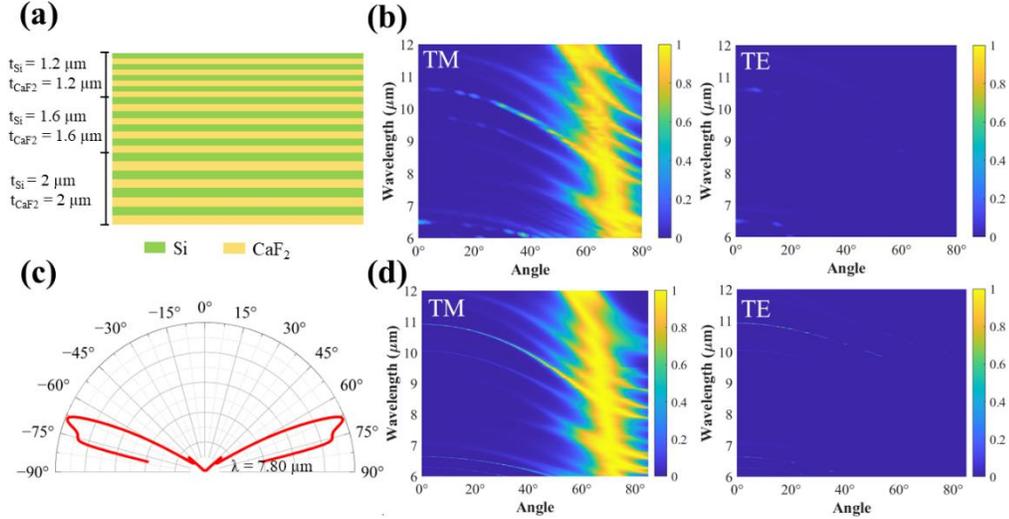

Fig. 6. a) Schematic of the multilayer filter structure embedded in a host material that is used for broadband directional thermal emission. b) Absorptance spectra of the FLSP surface covered by the multilayer structure and embedded in a host material for TM (left) and TE (right) polarization. These results were obtained through FEM simulations. c) Angular absorptance of the filter with FLSP at a fixed wavelength of $\lambda = 7.8$ μm. d) Transmittance spectra of the bare multilayer filter structure (i.e., without FLSP) embedded in a host material for TM (left) and TE (right) polarizations. These results were obtained analytically using the transfer matrix method.

In this design, the thermal radiation is restricted to an angular range between 60° and 80° over the entire mid-IR band from 6 μm to 12 μm. Since this design utilizes the Brewster angle effect, it only works for TM polarization, as it is obvious by inspecting Figs. 6b and 6d. Since the polarization of thermal radiation is random and the emissivity of the device is near zero for all angles and wavelengths only for TE polarization, the total emissivity values of this composite design will be essentially half of the presented values for TM polarization. However, this particular design has the additional useful functionality of operating as a polarizer filter for thermal emission, since it can emit only TM polarized broadband thermal radiation. This type of broadband directional thermal emission could be used for radiative cooling in cities [16] or as a coherent broadband directional antenna of IR light [69].

## 3. Conclusions

We have successfully demonstrated high performance daytime passive radiative cooling using a broadband FLSP thermal emitter covered with a solar Bragg visible reflector filter. This new passive cooling approach provides many benefits including the ability to place the thermal emitter in direct contact with the desired object that needs to be cooled, as opposed to more traditional radiative cooling designs using solar back reflectors [11,12,15,28–32]. The presented devices are expected to have useful applications in above-ambient temperature cooling and thermal management of aircraft or spacecraft. We have also demonstrated narrowband and broadband directional control of the thermal radiation from the presented FLSP metallic (aluminum-based) surfaces. These additional designs can be used as directional sources of IR light that can be used as new detectors and sensors. The ability to control and efficiently manipulate the spectral and directional characteristics of thermal radiation is a rapidly developing photonic engineering field which is expected to impact a wide variety of thermal management, sensing, and energy harvesting applications.

**Funding.** National Science Foundation (2224456); Office of Naval Research (N00014-19-1-2384 and N00014-20-1-2025). NASA Nebraska Space Grant.

**Disclosures.** The authors declare no conflicts of interest.

**Data availability.** Data underlying the results presented in this paper are not publicly available at this time but may be obtained from the authors upon reasonable request.

**References**

1. N. Picqué and T. W. Hänsch, "Mid-IR spectroscopic sensing," Opt. Photonics News 30(6), 26–33 (2019).
2. E. Sakr and P. Bermel, "Thermophotovoltaics with spectral and angular selective doped-oxide thermal emitters," Opt. Express 25(20), A880–A895 (2017).
3. R. Sakakibara, V. Stelmakh, W. R. Chan, M. Ghebrebrhan, J. D. Joannopoulos, M. Soljacic, and I. Čelanović, "Practical emitters for thermophotovoltaics: a review," J. Photon. Energy 9, 1 (2019).
4. A. LaPotin, K. L. Schulte, M. A. Steiner, K. Buznitsky, C. C. Kelsall, D. J. Friedman, E. J. Tervo, R. M. France, M. R. Young, A. Rohskopf, S. Verma, E. N. Wang, and A. Henry, "Thermophotovoltaic efficiency of 40%," Nature 604(7905), 287–291 (2022).
5. S. Jafari Ghalekohneh and B. Zhao, "Nonreciprocal Solar Thermophotovoltaics," Phys. Rev. Appl. 18(3), 034083 (2022).
6. D. Zhao, A. Aili, Y. Zhai, S. Xu, G. Tan, X. Yin, and R. Yang, "Radiative sky cooling: Fundamental principles, materials, and applications," Appl. Phys. Rev. 6(2), (2019).
7. J. W. Cho, Y. J. Lee, J. H. Kim, R. Hu, E. Lee, and S. K. Kim, "Directional Radiative Cooling via Exceptional Epsilon-Based Microcavities," ACS Nano 17(11), 10442–10451 (2023).
8. X. Liu, Y. Tian, F. Chen, A. Ghanekar, M. Antezza, and Y. Zheng, "Continuously variable emission for mechanical deformation induced radiative cooling," Commun. Mater. 1(1), 95 (2020).
9. K. Chen, M. Ono, S. Fan, and W. Li, "Self-adaptive radiative cooling based on phase change materials," Opt. Express 26(18), A777–A787 (2018).
10. Z. Huang and X. Ruan, "Nanoparticle embedded double-layer coating for daytime radiative cooling," Int. J. Heat Mass Transf. 104, 890–896 (2017).
11. A. P. Raman, M. A. Anoma, L. Zhu, E. Rephaeli, and S. Fan, "Passive radiative cooling below ambient air temperature under direct sunlight," Nature 515(7528), 540–544 (2014).
12. M. A. Kecebas, M. P. Menguc, A. Kosar, and K. Sendur, "Passive radiative cooling design with broadband optical thin-film filters," J. Quant. Spectrosc. Radiat. Transf. 198, 1339–1351 (2017).
13. E. Rephaeli, A. Raman, and S. Fan, "Ultrabroadband Photonic Structures to Achieve High-Performance Daytime Radiative Cooling," Nano Lett. 13(4), 1457–1461 (2013).
14. S. Y. Jeong, C. Y. Tso, Y. M. Wong, C. Y. H. Chao, and B. Huang, "Daytime passive radiative cooling by ultra emissive bio-inspired polymeric surface," Solar Energy Materials and Solar Cells 206, 110296 (2020).
15. A. K. Goyal and A. Kumar, "Recent advances and progresses in photonic devices for passive radiative cooling application: a review," J. Nanophotonics 14(03), 030901 (2020).
16. Y. Qu, M. Pan, and M. Qiu, "Directional and Spectral Control of Thermal Emission and Its Application in Radiative Cooling and Infrared Light Sources," Phys. Rev. Appl. 13(6), 064052 (2020).
17. D. G. Baranov, Y. Xiao, I. A. Nechepurenko, A. Krasnok, A. Alù, and M. A. Kats, "Nanophotonic engineering of far-field thermal emitters," Nat. Mater. 18(9), 920–930 (2019).
18. X. Li, J. Peoples, Z. Huang, Z. Zhao, J. Qiu, and X. Ruan, "Full Daytime Sub-ambient Radiative Cooling in Commercial-like Paints with High Figure of Merit," Cell Rep. Phys. Sci. 1(10), 100221 (2020).
19. Y. Zhai, Y. Ma, S. N. David, D. Zhao, R. Lou, G. Tan, R. Yang, and † Xiaobo Yin, "Scalable-manufactured randomized glass-polymer hybrid metamaterial for daytime radiative cooling," Science 355(6329), 1062–1066 (2017).
20. H. Bao, C. Yan, B. Wang, X. Fang, C. Y. Zhao, and X. Ruan, "Double-layer nanoparticle-based coatings for efficient terrestrial radiative cooling," Solar Energy Materials and Solar Cells 168, 78–84 (2017).
21. S. Atiganyanun, J. B. Plumley, S. J. Han, K. Hsu, J. Cytrynbaum, T. L. Peng, S. M. Han, and S. E. Han, "Effective Radiative Cooling by Paint-Format Microsphere-Based Photonic Random Media," ACS Photonics 5(4), 1181–1187 (2018).
22. J. Li, Y. Liang, W. Li, N. Xu, B. Zhu, Z. Wu, X. Wang, S. Fan, M. Wang, and J. Zhu, "Protecting ice from melting under sunlight via radiative cooling," Sci. Adv. **8**, 9756 (2022).

23. W. Zhu, B. Droguet, Q. Shen, Y. Zhang, T. G. Parton, X. Shan, R. M. Parker, M. F. L. De Volder, T. Deng, S. Vignolini, and T. Li, "Structurally Colored Radiative Cooling Cellulosic Films," Advanced Science **9**(26), (2022).
24. M. C. Huang, M. Yang, X. J. Guo, C. H. Xue, H. Di Wang, C. Q. Ma, Z. Bai, X. Zhou, Z. Wang, B. Y. Liu, Y. G. Wu, C. W. Qiu, C. Hou, and G. Tao, "Scalable multifunctional radiative cooling materials," Prog. Mater. Sci. **137**, (2023).
25. K. Lin, S. Chen, Y. Zeng, T. Chung Ho, Y. Zhu, X. Wang, F. Liu, B. Huang, C. Yu-Hang Chao, Z. Wang, and C. Yan Tso, "Hierarchically structured passive radiative cooling ceramic with high solar reflectivity," Science (1979) **382**, 691–697 (2023).
26. X. Zhao, T. Li, H. Xie, H. Liu, L. Wang, Y. Qu, S. C. Li, S. Liu, A. H. Brozena, Z. Yu, J. Srebric, and L. Hu, "A solution-processed radiative cooling glass," Science (1979) **382**(6671), 684–691 (2023).
27. J. Song, W. Zhang, Z. Sun, M. Pan, F. Tian, X. Li, M. Ye, and X. Deng, "Durable radiative cooling against environmental aging," Nat. Commun. **13**(1), (2022).
28. S. Y. Jeong, C. Y. Tso, J. Ha, Y. M. Wong, C. Y. H. Chao, B. Huang, and H. Qiu, "Field investigation of a photonic multi-layered TiO2 passive radiative cooler in sub-tropical climate," Renew. Energy 146, 44–55 (2020).
29. Y. Fu, J. Yang, Y. S. Su, W. Du, and Y. G. Ma, "Daytime passive radiative cooler using porous alumina," Solar Energy Materials and Solar Cells 191, 50–54 (2019).
30. Z. Chen, L. Zhu, A. Raman, and S. Fan, "Radiative cooling to deep sub-freezing temperatures through a 24-h day-night cycle," Nat. Commun. 7(1), 1–5 (2016).
31. J.-L. Kou, Z. Jurado, Z. Chen, S. Fan, and A. J. Minnich, "Daytime Radiative Cooling Using Near-Black Infrared Emitters," ACS Photonics 4(3), 626–630 (2017).
32. M. Lee, G. Kim, Y. Jung, K. R. Pyun, J. Lee, B. W. Kim, and S. H. Ko, "Photonic structures in radiative cooling," Light Sci. Appl. **12**(1), (2023).
33. L. Zhu, A. Raman, and S. Fan, "Color-preserving daytime radiative cooling," Appl. Phys. Lett 103(22), 223902 (2013).
34. D. Wu, C. Liu, Z. Xu, Y. Liu, Z. Yu, L. Yu, L. Chen, R. Li, R. Ma, and H. Ye, "The design of ultra-broadband selective near-perfect absorber based on photonic structures to achieve near-ideal daytime radiative cooling," Mater. Des. 139, 104–111 (2018).
35. B. Zhao, M. Hu, X. Ao, Q. Xuan, and G. Pei, "Comprehensive photonic approach for diurnal photovoltaic and nocturnal radiative cooling," Solar Energy Materials and Solar Cells 178, 266–272 (2018).
36. S. Campione, F. Marquier, J.-P. Hugonin, A. R. Ellis, J. F. Klem, M. B. Sinclair, and T. S. Luk, "Directional and monochromatic thermal emitter from epsilon-near-zero conditions in semiconductor hyperbolic metamaterials," Sci. Rep. 6(1), 1–9 (2016).
37. J. Xu, J. Mandal, and A. P. Raman, "Broadband directional control of thermal emission," Science (1979) 372, 393–397 (2021).
38. E. Sakat, G. Barbillon, J.-P. Hugonin, P. Ben-Abdallah, and S.-A. Biehs, "True thermal antenna with hyperbolic metamaterials," Opt. Express 25(19), 23356–23363 (2017).
39. B. J. Lee, C. J. Fu, and Z. M. Zhang, "Coherent thermal emission from one-dimensional photonic crystals," Appl. Phys. Lett. 87(7), 071904 (2005).
40. I. Celanovic, D. Perreault, and J. Kassakian, "Resonant-cavity enhanced thermal emission," Phys. Rev. B 72(7), 075127 (2005).
41. K. Ito and H. Iizuka, "Directional thermal emission control by coupling between guided mode resonances and tunable plasmons in multilayered graphene," J. Appl. Phys. 120(16), 163105 (2016).
42. A. R. Ellis, D. W. Peters, E. A. Shaner, P. S. Davids, and T. Ribaudo, "Highly directional thermal emission from two-dimensional silicon structures," Opt. Express 21(6), 6837–6844 (2013).
43. E. Sakr, S. Dhaka, and P. Bermel, "Asymmetric angular-selective thermal emission," in Physics, Simulation, and Photonic Engineering of Photovoltaic Devices V (International Society for Optics and Photonics, 2016), 97431D.
44. Q. Qian, C. Xu, and C. Wang, "All-dielectric polarization-independent optical angular filter," Scientific Report 7(1), 1–7 (2017).
45. Y. Shen, D. Ye, I. Celanovic, S. G. Johnson, J. D. Joannopoulos, and M. Soljačić, "Optical Broadband Angular Selectivity," Science (1979) 343(6178), 1499–1501 (2014).
46. A. Reicks, A. Tsubaki, M. Anderson, J. Wieseler, L. K. Khorashad, J. E. Shield, G. Gogos, D. Alexander, C. Argyropoulos, and C. Zuhlke, "Near-unity broadband omnidirectional emissivity via femtosecond laser surface processing," Commun. Mater. 2(1), 1–11 (2021).
47. L. K. Khorashad, A. Reicks, A. Erickson, J. E. Shield, D. Alexander, A. Laraoui, G. Gogos, C. Zuhlke, and C. Argyropoulos, "Unraveling the formation dynamics of metallic femtosecond laser induced periodic surface structures," Opt. Laser Technol. 171, (2024).
48. A. T. Tsubaki, M. A. Koten, M. J. Lucis, C. Zuhlke, N. Ianno, J. E. Shield, and D. R. Alexander, "Formation of aggregated nanoparticle spheres through femtosecond laser surface processing," Appl. Surf. Sci. **419**, 778–787 (2017).
49. E. Peng, R. Bell, C. A. Zuhlke, M. Wang, D. R. Alexander, G. Gogos, and J. E. Shield, "Growth mechanisms of multiscale, mound-like surface structures on titanium by femtosecond laser processing," J. Appl. Phys. **122**(13), (2017).
50. L. K. Khorashad, A. Reicks, A. Erickson, J. E. Shield, D. Alexander, A. Laraoui, G. Gogos, C. Zuhlke, and C. Argyropoulos, "Unraveling the formation dynamics of metallic femtosecond laser induced periodic surface structures," Opt. Laser Technol. **171**, (2024).

51. A. D. Rakić, "Algorithm for the determination of intrinsic optical constants of metal films: application to aluminum," Appl. Opt. 34(22), 4755 (1995).
52. J. Kischkat, S. Peters, B. Gruska, M. Semtsiv, M. Chashnikova, M. Klinkmüller, O. Fedosenko, S. MacHulik, A. Aleksandrova, G. Monastyrskyi, Y. Flores, and W. T. Masselink, "Mid-infrared optical properties of thin films of aluminum oxide, titanium dioxide, silicon dioxide, aluminum nitride, and silicon nitride," Appl. Opt. 51(28), 6789–6798 (2012).
53. M. R. Querry, "Optical constants," Contractor Report 415, 1001 (1985).
54. A. Y. Vorobyev, V. S. Makin, and C. Guo, "Brighter light sources from black metal: Significant increase in emission efficiency of incandescent light sources," Phys. Rev. Lett. 102(23), (2009).
55. "Reference Air Mass 1.5 Spectra | Grid Modernization | NREL," https://www.nrel.gov/grid/solar-resource/spectra-am1.5.html.
56. D. Pozar, Microwave Engineering, 1st ed. (Wiley, 1998).
57. T. Siefke, S. Kroker, K. Pfeiffer, O. Puffky, K. Dietrich, D. Franta, I. Ohlídal, A. Szeghalmi, E. B. Kley, and A. Tünnermann, "Materials Pushing the Application Limits of Wire Grid Polarizers further into the Deep Ultraviolet Spectral Range," Adv. Opt. Mater. 4(11), 1780–1786 (2016).
58. J. Kischkat, S. Peters, B. Gruska, M. Semtsiv, M. Chashnikova, M. Klinkmüller, O. Fedosenko, S. MacHulik, A. Aleksandrova, G. Monastyrskyi, Y. Flores, and W. T. Masselink, "Mid-infrared optical properties of thin films of aluminum oxide, titanium dioxide, silicon dioxide, aluminum nitride, and silicon nitride," Appl. Opt. 51(28), 6789–6798 (2012).
59. S. Popova, T. Tolstykh, and V. Vorobev, "Optical characteristics of amorphous quartz in the 1400–200 cm-1 region," Opt. Spectrosc. 33, 444–445 (1972).
60. H. H. Li, "Refractive index of silicon and germanium and its wavelength and temperature derivatives," J. Phys. Chem. Ref. Data 9(3), 561–658 (1980).
61. C. Schinke, P. Christian Peest, J. Schmidt, R. Brendel, K. Bothe, M. R. Vogt, I. Kröger, S. Winter, A. Schirmacher, S. Lim, H. T. Nguyen, and D. Macdonald, "Uncertainty analysis for the coefficient of band-to-band absorption of crystalline silicon," AIP Adv. 5(6), (2015).
62. C. G. Granqvist and A. Hjortsberg, "Radiative cooling to low temperatures: General considerations and application to selectively emitting SiO films," J. Appl. Phys. 52(6), 4205–4220 (1981).
63. "ATRAN," https://atran.arc.nasa.gov/cgi-bin/atran/atran.cgi.
64. Gemini Observatory, "IR Transmission Spectra," https://www.gemini.edu/observing/telescopes-and-sites/sites#Transmission.
65. S. D. Lord, "A new software tool for computing Earth's atmospheric transmission of near- and far-infrared radiation," NASA Technical Memorandum 103957 (1992).
66. A. Shitzer, "Wind-chill-equivalent temperatures: regarding the impact due to the variability of the environmental convective heat transfer coefficient," Int. J. Biometeorol. 50(4), 224–232 (2006).
67. H. H. Li, "Refractive index of alkaline earth halides and its wavelength and temperature derivatives," J. Phys. Chem. Ref. Data 9, 161–290 (1980).
68. A. Butler, J. Schulz, and C. Argyropoulos, "Tunable directional filter for mid-infrared optical transmission switching," Opt. Express 30(22), 39716 (2022).
69. C. Argyropoulos, K. Q. Le, N. Mattiucci, G. D'Aguanno, and A. Alù, "Broadband Absorbers and Selective Emitters based on Plasmonic Brewster Metasurfaces", Phys. Rev. B 87(20), 205112 (2013).